\documentclass[preprint,showkeys,notfootinbib]{revtex4}
\usepackage[utf8]{inputenc}
\usepackage{amssymb}
\usepackage{color}
\usepackage{amsmath}
\usepackage[font={small}]{caption}
\usepackage{epsf}
\usepackage{psfrag}
\usepackage{graphicx}
\usepackage{caption}
\usepackage{epstopdf}

\newcommand{\ba}{\begin{align}}
\newcommand{\ea}{\end{align}}
\newcommand{\bq}{\begin{equation}}
\newcommand{\eq}{\end{equation}}
\newcommand{\barr}{\begin{eqnarray}}
\newcommand{\earr}{\end{eqnarray}}

\begin{document}

\title{Unifying turnaround-bounce cosmology in a cyclic universe considering a running vacuum model}
\author{ A. Khodam-Mohammadi
\footnote{Email:\text{khodam@basu.ac.ir}}}
\affiliation{Department of Physics, Faculty of Science, Bu-Ali Sina
University, Hamedan 65178, Iran}

\begin{abstract}
 Among many models which can describe the bouncing cosmology, A matter bounce scenario that is deformed by a running vacuum model of dark energy (RVM-DE) has been interested. In this research, I show that a class of RVM-CDM (cold dark matter) model can also describe a cyclical cosmology in which the universe undergoes cycles of expansion to the contraction phase and vice versa. To this end, following our previous work, I consider one of the most successful class of RVM-CDM model in bouncing cosmology, $\rho_x=n_0+n_2 H^2 +n_4 H^4$, in which the power spectral index gets a red tilt and the running of the spectral index may give a negative value by choosing the appropriate value of parameters ($n_0,~n_2,~n_4$), which is consistent with the cosmological observations. It is worthwhile to mention that most matter bounce models do not produce this negative value. However, the main purpose of this article is to investigate the RVM-CDM model in the turnaround phase. Far from the bounce in a phantom expanding universe, the turnaround conditions are investigated before the occurrence of a sudden big rip. By analyzing the Hubble parameter, equation of state parameter, and deceleration parameter around the turnaround, we show that a successful turnaround may occur after an expansion in an \textbf{interacting} case of RVM-CDM by choosing the appropriate value of parameters. A minimum value for the interaction parameter is obtained and also find any relation between other model parameters. Finally, the effect of each parameter on a turnaround is studied, and we see that the transition time from accelerating to decelerating expansion can occur earlier for larger values of interaction parameter. Also, in several graphs, the effect of the second term in DE density, including	$H^2$, is studied, and we see that by increasing its coefficient, $n_2$, the transition point leads to lower values. 
\end{abstract}

%\PACS{{95.36.+x}{Dark energy} \and{98.80.-k}{Cosmology}}

\maketitle

\section{introduction}

For centuries, the debate between a cyclical eternal universe and an accidental one, has been the subject of controversy between philosophers and scientists.
 In the twentieth century, after the introduction of Einstein's theory of gravity, Tolman \cite{Tolman_1934} returned to the hypothesis of a cyclic universe by using a positive spatially curvature cosmology. Since then, many scientists have proposed different models for this type of universe.
These models were presented to resolve the problem of big bang
theory, which has an avoidable initial singularity of the universe.
 Furthermore, in many of cosmological dark energy (DE) models, the existence of a far future singularity,
 called the big rip, has been predicted \cite{Caldwell:2003vq,Nojiri:2011kd}. A cyclic universe must re-collapse before
 the big rip occurs. The fundamental question is, what physical process causes these
 cycles?

One of the interesting models of cyclic cosmology to date is the \emph{Conformal Cyclic Cosmology} ( CCC) model, which was
popularized by a Nobel laureate, Roger Penrose, in 2010 in the book:
\emph{The Cycle of Time} \cite{Penrose2011CyclesOT}. Penrose group
and other cosmologists have written many articles on the evidence of
CCC in observations of the cosmic microwave background (CMB) and its
anomalies
\cite{Penrose:2006zz,Penrose:2010zz,Gurzadyan:2011ac,Gurzadyan:2013cna,Gurzadyan:2015gzb,Penrose:2017npq,An:2018utx}.

Other successful models that have been also considered in cyclic
cosmology are: Steinhardt Turok model
\cite{Steinhardt:2001st,Steinhardt:2002ih,Steinhardt:2002kw,Boyle:2003km,Khoury:2003rt,Steinhardt:2004gk,
Lehners:2008my,Lehners:2013cka,Ijjas:2016tpn,Ijjas:2016vtq,Ijjas:2018qbo,Ijjas:2021zwv,Ijjas:2019pyf},
Baum-Frampton model
\cite{Baum:2006ee,Frampton:2006mg,Frampton:2007cv,Arefeva:2008abt,Frampton:2013fxa,Coriano:2019eif},
and Loop quantum cosmology (LQC).

In LQC model, physicists approach the concepts of quantum mechanics
 and quantum gravity. A non-perturbative regime of quantum gravity  based on loop
 quantum gravity (LQG), may stand out to arise a nonsingular big bounce or big crunch
 once before tending to singularity, by some matter-energy content or spatial curvature,
 for various isotropic and anisotropic models \cite{Singh:2009mz}. Following the LQG,
 the cosmologists introduce an LQC theory, based on discrete quantum geometry
 arising from the non-perturbative quantum geometric effects
 \cite{Ashtekar:2011ni}. Using LQC, the Friedmann equation is modified by a quadratic of energy
 density term (with a negative sign), which in turned into standard Friedmann equation,
 a little far from the big bounce and turnaround points \cite{WilsonEwing:2012pu}.
 The Hubble rate is vanished and changed the sign at maximum energy in a
 finite value of scale factor. The relic gravitational waves
 from the merging of a large number of black holes in a collapsed phase
 of the universe in the pre-bounce regime, may be regarded as good evidence of a cyclic universe \cite{Gorkavyi:2021tbw}.

Among many models of bouncing cosmology, i.e. Pre-Big-Bang
\cite{Gasperini:1992em} or Ekpyrotic type \cite{Khoury:2001wf},
string gas cosmology \cite{Brandenberger:1988aj, Nayeri:2005ck} and
matter bounce scenario \cite{Finelli:2001sr}, the last one has the
most interested in the last decade
\cite{Lin:2010pf,WilsonEwing:2012pu,Cai:2013kja,Lehners:2015mra,Cai:2015vzv,Cai:2016hea}.
 Based on observations of Plank 2015
and 2018
\cite{Planck:2015sxf,Planck:2015fie,Ade:2015tva,BICEP2:2018kqh}, the
primordial perturbations were adiabatic, almost scale-invariant with
a slight red tilt ($n_s=0.968\pm 0.006,~ (68\%~ CL)$), and that
tensor perturbations were small. Also these observations provide a
bound for running of spectral index $\alpha_s=-0.003\pm 0.007,~
(68\%~ CL)$. In particular, some efforts has been made in a
 quasi-matter, semi-matter, and
deformed matter (a matter term mixed by a dark energy component)
bouncing cosmology \cite{Cai:2014zga,Cai:2014jla,Arab:2017gae},
where in some of which, the problems of previous models have been
solved. Although in many models of matter bounce, the power spectral
index ($n_s$) of cosmological perturbation may be consistent with
observations, the sign of running of the spectral index ($\alpha_s$)
is one of the matter of challenges among the models.
 In fact, the quantity $\alpha_s$ may
be considered as a further observational tool that can be used to
differentiate between various cosmological scenarios
\cite{Lehners:2015mra}. Most recently, in \cite{Arab:2021wis},
authors made constraint parameters of a quasi-matter bounce model in
light of Plank and BICEP2/Keck data set.

In this paper, following our previous work \cite{Arab:2017gae}, in
which a successful model of deformed-matter bounce (a pressureless
matter deformed by a running vacuum model (RVM) of DE, with a small
tiny negative value of the equation of state parameter, $w$) was
introduced, now I am also interested to use this model in a cyclic
cosmology, especially in the turnaround point.

 In recent years, several models have
been proposed in the field of cyclic cosmology, but the complete
cycle from bounce to turnaround has not been studied in any of them.
\cite{Zhang:2007an,Sheykhi:2017cid}.

Before getting started, it must be noted that at following, I am using the reduced Planck mass unit system, in which $\hbar=c=8\pi G=1$.

\section{Cyclic cosmology in a RVM-CDM universe }
 In a flat FLRW universe,
\bq
 ds^2=-dt^2+a(t)^2[dr^2+r^2 d\theta ^2+r^2 sin^2(\theta) d\phi ^2], \label{4}
\eq
which is fulfilled by three components, dark energy (DE), cold dark matter (CDM)
and radiation, a holonomy corrected Loop Quantum Cosmology (LQC) in high energy context
of cosmology, gives approximately full quantum dynamics of the universe by
introducing the following set of effective equations \cite{Ashtekar:2006wn}
\barr
H^2&=&\frac {\rho}{3}(1-\frac{\rho}{\rho_c}),
\label{3}\\
\dot H&=&(\frac{1}{2} \rho-3H^2)(1+w), \label{5}
 \earr
 where
$ \rho= \rho_m+\rho_r+\rho_x $ is the total energy density inclusive
of pressureless CDM, radiation and DE respectively. The quantity
$\rho_c$ is critical energy density which is around the Planck
energy density ($\rho_c\sim\rho_{Pl}$). Also one can easily see that
by $ \rho_c \to \infty $, the classical Friedmann equations in the
flat universe are retrieved. Same as low energy cosmology, in this
context, the continuity equation easily obtained
 \bq
 \dot{\rho}+3H\rho(1+w)=0, \label{4}
\eq
  where $w=P/\rho$ is the  effective equation of state parameter. The continuity equation (\ref{4}) can be decomposed by three
 equations for all components of energy as
 \barr
\dot{\rho_x}&=&-Q\notag\\
   \dot{\rho}_m+3H\rho_m&=& Q \label{1}\\
 \dot\rho _r+4H\rho_r&=&0\notag
  \earr
 where $Q$ plays an interaction term between DE and CDM
 and superscript dot refers to derivative with respect to cosmic time.
 We must mentioned that in this paper, we are interested to use of the
 running vacuum model (RVM) in which the equation of state parameter
 (EoS) is '$w_x=-1$', like as rigid cosmological constant $\Lambda$
 \cite{Perico:2013mna}.
  The running vacuum energy density is not a constant but rather it
is a function of the cosmic time. Note that in many papers it has
been expanded as a function of the Hubble parameter. The nature of
RVM is essentially connected with the renormalization group (RG) in
quantum field theory (QFT) in curved spacetime. In this context, the
evolution of the vacuum is written as a function of $H^2$ which
determines the running of the vacuum energy
\cite{Shapiro:1999zt,Shapiro:2000dz,Babic:2004ev,Shapiro:2003ui,Espana-Bonet:2003qjh,
Shapiro:2003kv,Sola:2007sv,Shapiro:2009dh,Basilakos:2009wi,Grande:2011xf,Basilakos:2012ra,Lima:2013dmf}.
 In any expansion-contraction transition time (bounce/turnaround),
 the Hubble parameter evaluates to be vanishes and energy density mimics
 high energy Planck density, which is supplied by radiation term in
 big bounce case and by DE term in turnaround. By  generalizing a special
 case of RVM in an XCDM flat universe  (a dynamical dark energy-CDM), the possibility of growth of DE
 density up to Planck energy density will be realized. The energy density
 of  Dynamical-DE can be considered as
 a driving  energy of LQC to originate a turnaround.
  However, at last, I show that at the turnaround point the
 dynamical DE will be merged to RVM in which $w_x=-1$, which has been used around the bouncing.
 \section{Bouncing phase in RVM-CDM universe }
 In this section, I give a brief review of
the bouncing cosmology specially in deformed matter bounce
scenario \cite{Cai:2014jla,Arab:2017gae}. \\
Generally as explained in the previous section, in QFT, a
theoretical
 explanation of the RVM-DE is given by \cite{Arab:2017gae,
 Sola:2013gha,Sola:2015rra,Sola:2016jky}
  \bq
   \rho_x=\rho_{\Lambda} (H^2,\dot H)=n_0+n_2 H^2+\alpha\dot H+n_4 H^4 +O(H^6).\label{varlambda}
   \eq
  Note that the first term ($n_0$) in (\ref{varlambda}), has the role of standard
 rigid $\Lambda$ model \cite{Cai:2014jla}. In a RVM-CDM bouncing
 scenario, By solving effective equations (\ref{3}, \ref{5}, \ref{4}), one can
 find the behavior of scale factor, EoS parameter, and Hubble rate
 in some cases of RVM-CDM as well as $\Lambda$CDM model
 in the background, around the bouncing point \cite{Cai:2014jla,Arab:2017gae}.
 It should be noted that in the bounce time ($t=0$), the Hubble
parameter should vanish, and the radiation term becomes a dominant
term in the total energy density. In this phase, the scale factor
becomes a non-zero minimum value where we can normalize its value to
unity.\\
Up to this level, in the background examination, almost all matter
and deformed matter bounce scenarios, are acceptable. Their
differences will appear in the analysis of models under the theory
of cosmological perturbation.\\
The linear primordial perturbation extended into LQC has been
studied by effective Mukhanove-Sasaki equation
\cite{Wilson-Ewing:2011gnq,Cailleteau:2012fy}. In this theory, the
scalar spectral index of power spectrum and its running, at the
crossing time, which is a time before the bounce when the sound
horizon crossed by long wavelength modes ($k=a|H|$) and it gives a
good condition to solve the perturbation equation, was given by
\cite{Arab:2017gae},
  \bq
     n_s=1+12(w_0+\frac{\alpha_{sc}}{12})\approx 1+12w,
     \label{spectral}
     \eq.
 \bq
   \alpha_{sc}= \left( \frac{ d n_s}{ d \ln k}\right)_{k=a|H|
   }= \frac{12H \dot{w}}{H^2+\dot{H}}\cong3 H \frac{d n_s}{d H},\label{alphass}
     \eq
 where $\alpha_{sc}$ is the running
of spectral index and $w_0$ is the EoS parameter at the crossing
time. Note that the effective equation of state parameter has a tiny
non-constant negative value in the contracting phase of the universe
at the crossing time. In a $\Lambda$CDM bounce scenario
\cite{Cai:2014jla}, although the scalar spectral index $n_s$ gets a
red tilt, the running of spectral index gets a positive value (see
the first row of table \ref{chartt}). In another case, row 5 of
table \ref{chartt}, the EoS parameter is constant which results a
constant $n_s$ and finally gets $\alpha_{sc}=0$. In some cases, rows
2-4 of table \ref{chartt}, we obtain $\alpha_{sc}>0$ and only in one
case of RVM-CDM, which I studied (row 6), we see $\alpha_{sc}<0$.
In table \ref{chartt}, the effective equation of state parameter and
running $\alpha_{sc}$ at crossing time for some cases in RVM-CDM
bounce scenario have been calculated \cite{Arab:2017gae}.
 \begin{table*}[h]
\begin{center}
\begin{tabular}{|r|c|c|c|l|}
\hline

  &$\rho_{\Lambda}$ & $ w $ & $\alpha_s$ & $\alpha_{sc}$\\
\hline
 1 &$n_0$ & $\dfrac{-n_0}{3H^2}$ & $-72 w$ & 0.22\\
\hline
  2 &$n_0+n_2 H^2$ & $ -\dfrac{1}{3}(\dfrac{n_0}{H^2}+n_2)$ & $\dfrac{n_0}{n0+n_2H^2}(-72 w) $ & 0.22 \\
\hline
 3 &$n_0+n_1 H$ &$-\dfrac{1}{3}(\dfrac{n_0}{ H^2}+\dfrac{n_1}{ H})$ & $\dfrac{n_1 H+2 n_0}{n_1 H+n_0}(-72 w)$ & 0.44\\

\hline

 4 &$n_0+\gamma \dot{H}$ &$ \dfrac{2 n_0/H^2-3 \gamma}{-6+3\gamma}$ & $\dfrac{n_0}{n_0-3 \gamma H^2 /2} (-72 w)$ & 0.001\\

\hline

 5 &$n_2 H^2 +\gamma \dot{H}$&$ \dfrac{-2 n_2+3 \gamma}{ 6-3 \gamma}$ &0& 0\\
\hline

 6 &$n_0+n_2 H^2 +n_4 H^4$ &$ -\dfrac{1}{3}(\dfrac{n_0}{H^2}+n_2+n_4 H^2)$ & $ \dfrac{(-n_4 H^4+n_0)(-72 w)}{n_0+n_2 H^2 +n_4 H^4}$ & -0.003\\

\hline
\end{tabular}
\end{center}
\caption{Summary of all cases in RVM-bounce scenario at crossing
time ($H_{cr}=-8.8* 10^{-8}, ~w_{cr}=-0.003 $) \cite{Arab:2017gae}.
 }
\label{chartt}
\end{table*}
 As we see in the last row of table \ref{chartt}, the running of the spectral index has a negative value
 and by considering preferred values of parameters, this model gives the best consistency with the
 cosmological observations (based on 2013 and 2015 Plank results (Ade et al. 2014, 2016),
 the running was provided a tiny negative value) \cite{Arab:2017gae}.

  It is worthwhile to mention that unlike matter bounce scenario
 which gives a positive value of the running
 of spectral index \cite{Lehners:2015mra}, in the deformed matter bounce scenario,
 there is a possibility of the negative running of spectral
 index and this can be the strength of this type of scenario.
 \section{turnaround in a RVM-CDM universe }
 Regardless of any particular model in cosmology, the turnaround point must have the following characteristics.
 First, the scale factor would reach a finite maximum value, in which the Hubble parameter
 will vanish at this point. In fact, the Hubble parameter will change from a positive value to a negative value.
 Second, the condition of reaching the high energy phase and supplying the LQC must be met. Our hypothesis is that just as at the time of bounce, where the energy density of radiation causes a critical density to reach the LQC energy level, in the turnaround phase, The dark energy that grows up to the critical energy, provides LQC energy condition. To prove this claim, we need a model of dark energy whose energy density can grow in the phantom phase to reach the critical energy and finally remains at the phantom wall in the rapid contraction phase of the universe. Therefore, the universe must remain in the phantom phase in an expanding regime.
 Also, the sign of deceleration parameter $q$, must change from a negative to a positive value.
 Third, due to a destructive effect of an expanding universe in the phantom phase in the creation
 of a sudden big rip,  the turnaround point must be realized before the big rip occurs.

 Now at following, we study the successful case, row 6  of table \ref{chartt},
 in the RVM-CDM model, around the turning point, by considering a universe with and without any terms of interaction.

\subsection{debate of interacting and non-interacting case}
In an interacting dynamical DE model, in which the equation of state of dark sector is a function of cosmic time, the continuity equations are
 \barr
\dot{\rho_x}+3 H \rho_x (1+w_x)&=&-Q\label{eqint1}\\
   \dot{\rho}_m+3H\rho_m&=& Q ,
 \earr
where $Q$ is an interaction term between dark matter (DM)-DE. By giving {\cite{Sheykhi:2017cid}}
  \bq
  Q =3 b^2 H (\rho_x +\rho_m) \label{eqint2}\\
    \eq
 and differentiating of corrected Friedmann equation (\ref{3}), we find
\bq
\dot{H}=- \frac{\rho_x}{2} (1+w_x +u)(1-\frac{2 \rho}{\rho_c}). \label{Hd}
\eq
 In the above equation, the quantity $u=\rho_m /\rho_x$ is the ratio of energy densities of DM to DE.

 After defining the dimensionless dark energy density
 \bq
 \Omega_x=\frac{\rho_x}{3 H^2}, \label{Omx1}
 \eq
 and using
 \bq
 1-\frac{2 \rho}{\rho_c}=\frac{2-\Omega_x (1+u)}{\Omega_x (1+u)},
 \eq
 which is derived from Eq. (\ref{3}), the previous Eq. (\ref{Hd}) yields
 \bq
 \frac{\dot{H}}{H^2}=-\frac{3}{2}(1+w_x+u)\frac{2-\Omega_x (1+u)}{(1+u)}. \label{HdH2}
 \eq
 Near the turning point, where the DE is the only dominated term, from (\ref{3}), we find
 \bq
 \Omega_x \approx 1+\frac{\rho_x}{\rho_c -\rho_x}, \label{Omx2}
\eq
and by defining a new relative parameter $\theta=\rho_x/\rho_c$, the Friedmann equation at the high energy level ($\rho\approx\rho_x$), can be rewritten as
\bq
 \Omega_x \approx \frac{1}{1-\theta}. \label{Omx3}
\eq
Regarding the new parameter, it is worth noting that due to the temporal increase of $ \rho_x $ in a
phantom universe, the parameter $ \theta $ also increases with time. Therefore, the study of other
parameters in terms of $ \theta $ can be equivalent to the time consideration of those parameters.
Eequation (\ref{Omx3}) shows that in a flat cyclic universe unlike standard cosmology, the
quantity $\Omega_x$ always greater than unity, so that in the limiting case, far from the turning point,
where $\rho_x \ll \rho_c$, it gives $\Omega_x \rightarrow 1^+$ and at the turning point,
where $\rho \approx\rho_x \approx \rho_c$ and $u\approx 0$, it gives $\Omega_x \rightarrow \infty$.

 Using the last model of table \ref{chartt}, which has the most consistency with the cosmological data
 \bq
 \rho_x=n_0+n_2 H^2+n_4 H^4. \label{rox1}
\eq
 Its time derivative gives
 \bq
 \dot{\rho_x}=4H \rho_x (1-\epsilon -\frac{n_2}{6 \Omega_x})(\frac{\dot{H}}{H^2}),\label{rhod2}
 \eq
 where the parameter $\epsilon=n_0/\rho_x$ is a new parameter, function of $\theta$.

 In order to examine the Hubble parameter, considering the Eq. (\ref{rhod2}) around
 the turning point ($u\approx 0, ~ \Omega_x\approx 1/(1-\theta)$),
 using $\dot{\theta}=\dot{\rho_x}/\rho_c$ and expanding the
 function $\epsilon (\theta)\approx\epsilon_0+(\theta-1)\epsilon _1$
 around the turning point $\theta=1$ up to first order, we obtain
 \bq
 \frac{\dot{\theta}}{H}\approx \dfrac{2\theta}{3}[(6\epsilon_1 -n_2)(1-\theta)-6\epsilon_0 +6](\dfrac{\dot{H}}{H^2}).
 \eq
 It is worth mentioning that the constant $ \epsilon_0(\theta =1) $ must be unity
 because at the turning point we have  $\rho=\rho_x=\rho_c=n_0$ since $H=0$. Now using
 \bq
 \frac{1}{H}\frac{d{H}}{d\theta}= \frac{\dot{H}/H^2}{\dot{\theta}/H},
 \eq
 we have
 \bq
 \frac{1}{H} \frac{dH}{d\theta}\approx \frac{3}{2\theta}[(6\epsilon_1 -n_2)(1-\theta)]^{-1}. \label{calcH1}
 \eq
 By integrating (\ref{calcH1}), we obtain
 \bq
 H=H_0\Bigg[\frac{(n_2-6\epsilon_1)(1-\theta)}{\theta}\Bigg]^{(\frac{3}{2 n_2-12 \epsilon_1})},\label{H1}
 \eq
 where the quantity $H_0$ is the constant of integration. A closer look at this
 equation shows that in order to avoid divergent $ H $ at $ \theta=1 $, we must have $ n_2 \geq 6\epsilon_1 $.

\begin{figure}[!htb]
    \begin{minipage}{0.48\textwidth}
        \centering
        \includegraphics[scale=0.42]{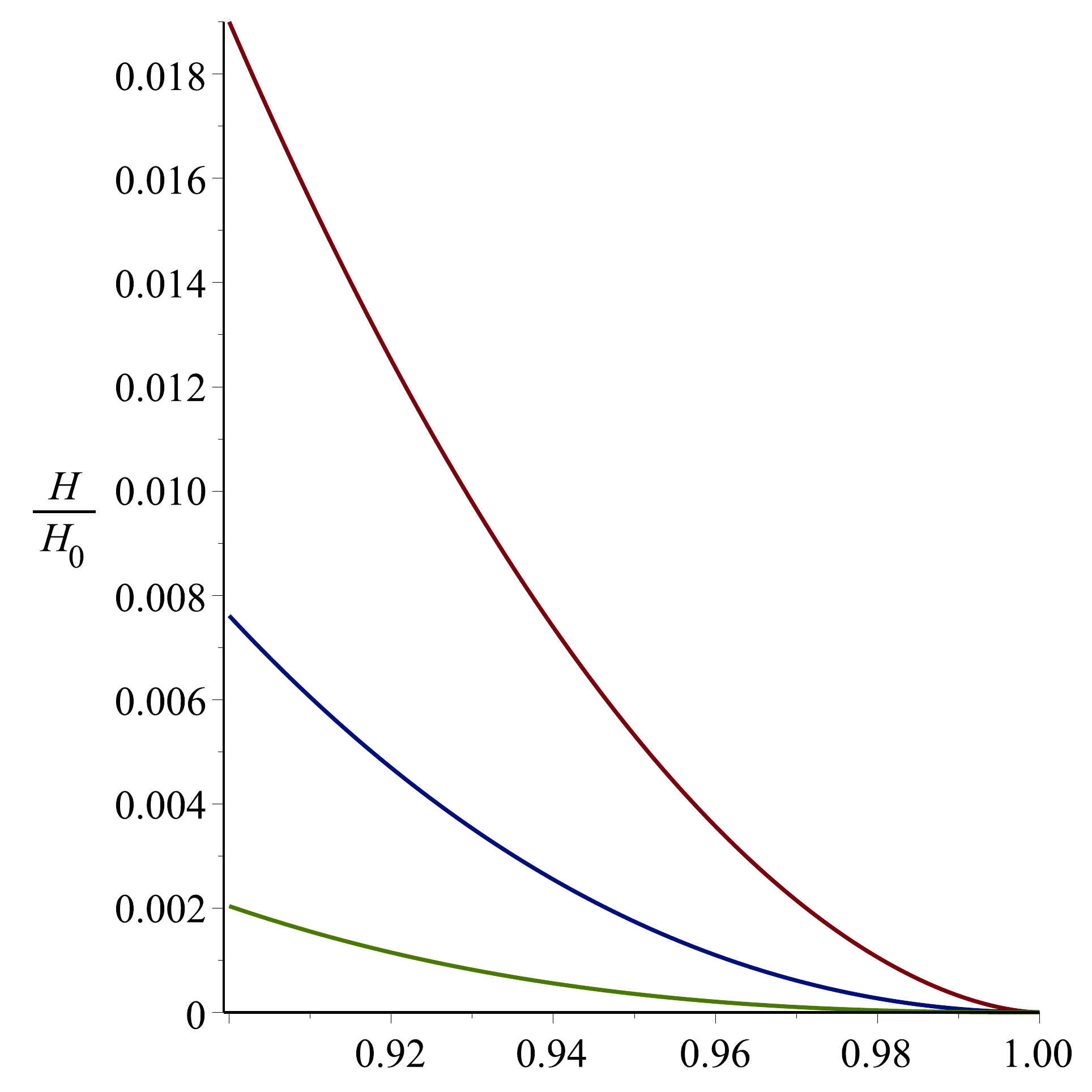}
        \caption{\small{The evolution of $\frac{H}{H_0}$ versus $\theta$ for various $\epsilon_1=0.02,~ 0.04,~ 0.06 $ (red, blue, green)}} \label{He}
    \end{minipage}\hfill
    \begin{minipage}{0.48\textwidth}
        \centering
        \includegraphics[scale=0.42]{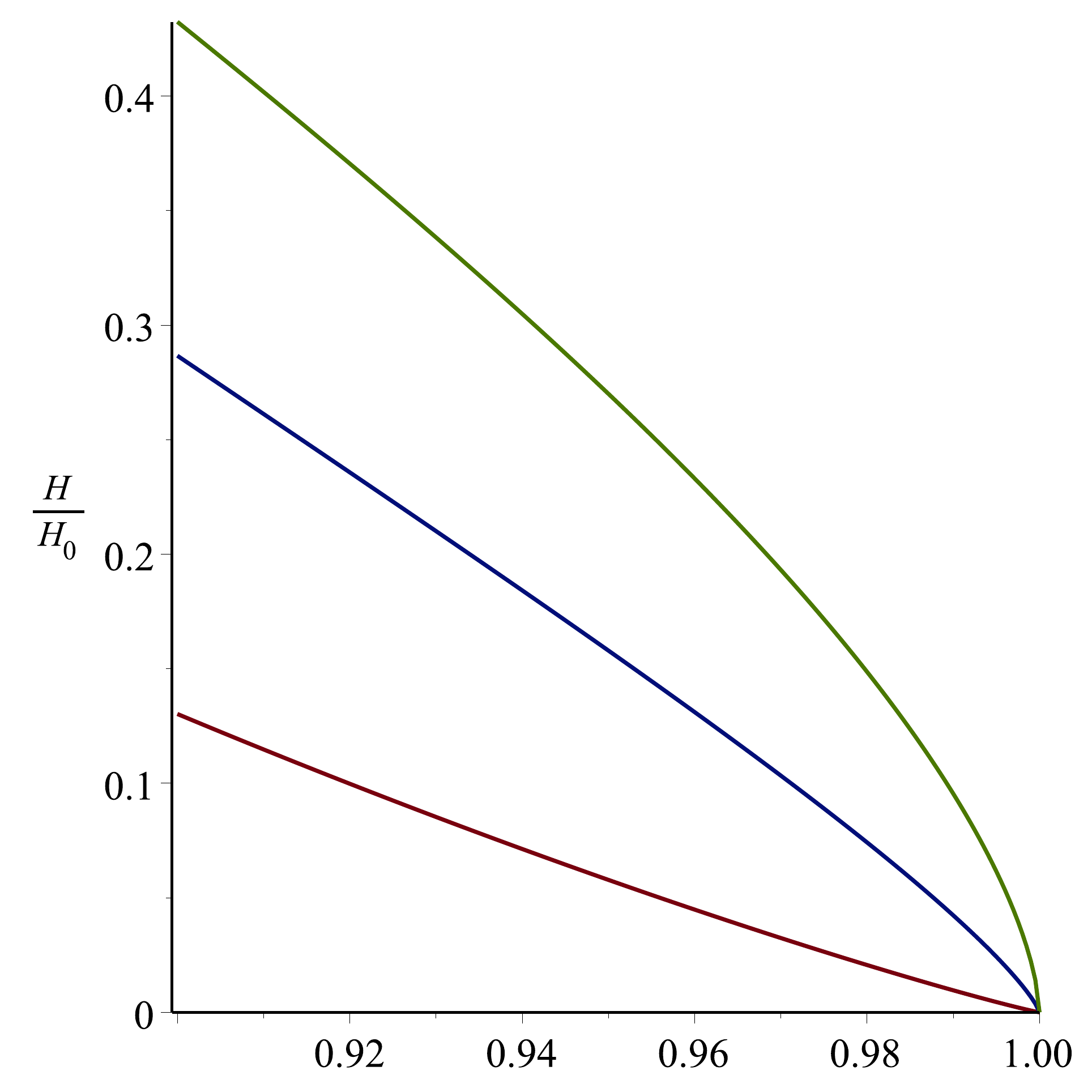}
        \caption{\small{The evolution of $\frac{H}{H_0}$ versus $\theta$ for various $n_2=1.5,~ 2,~ 2.5 $ (red, blue, green)}} \label{Hn}
    \end{minipage}
 \end{figure}

\begin{figure}[!htb]
    \begin{minipage}{0.48\textwidth}
        \centering
    \includegraphics[scale=0.42]{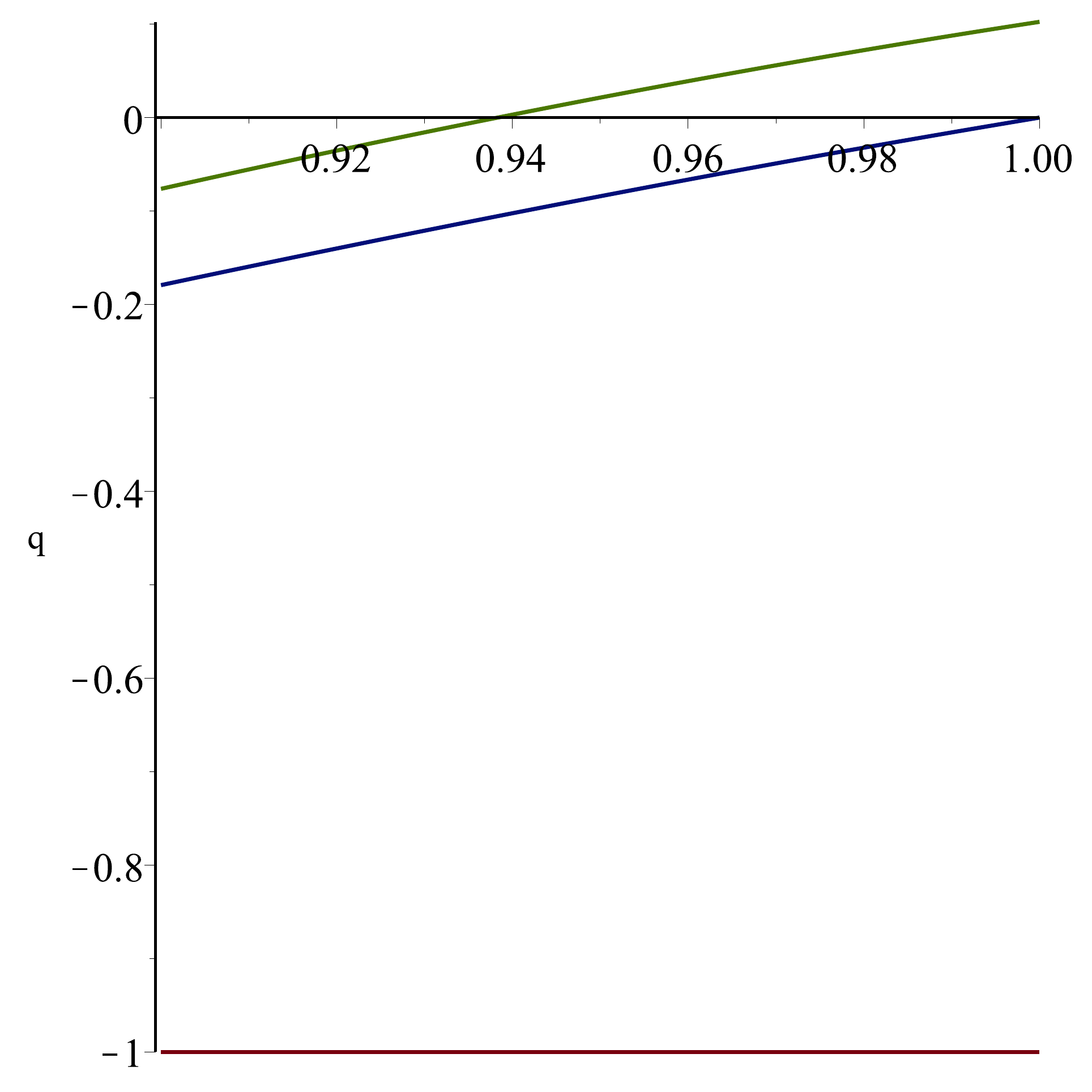}
        \caption{\small{The evolution of $q$ versus $\theta$ for various $b=0,~ 2/3,~ 0.7 $ (red, blue, green)}} \label{qb}
    \end{minipage}\hfill
    \begin{minipage}{0.48\textwidth}
        \centering
        \includegraphics[scale=0.42]{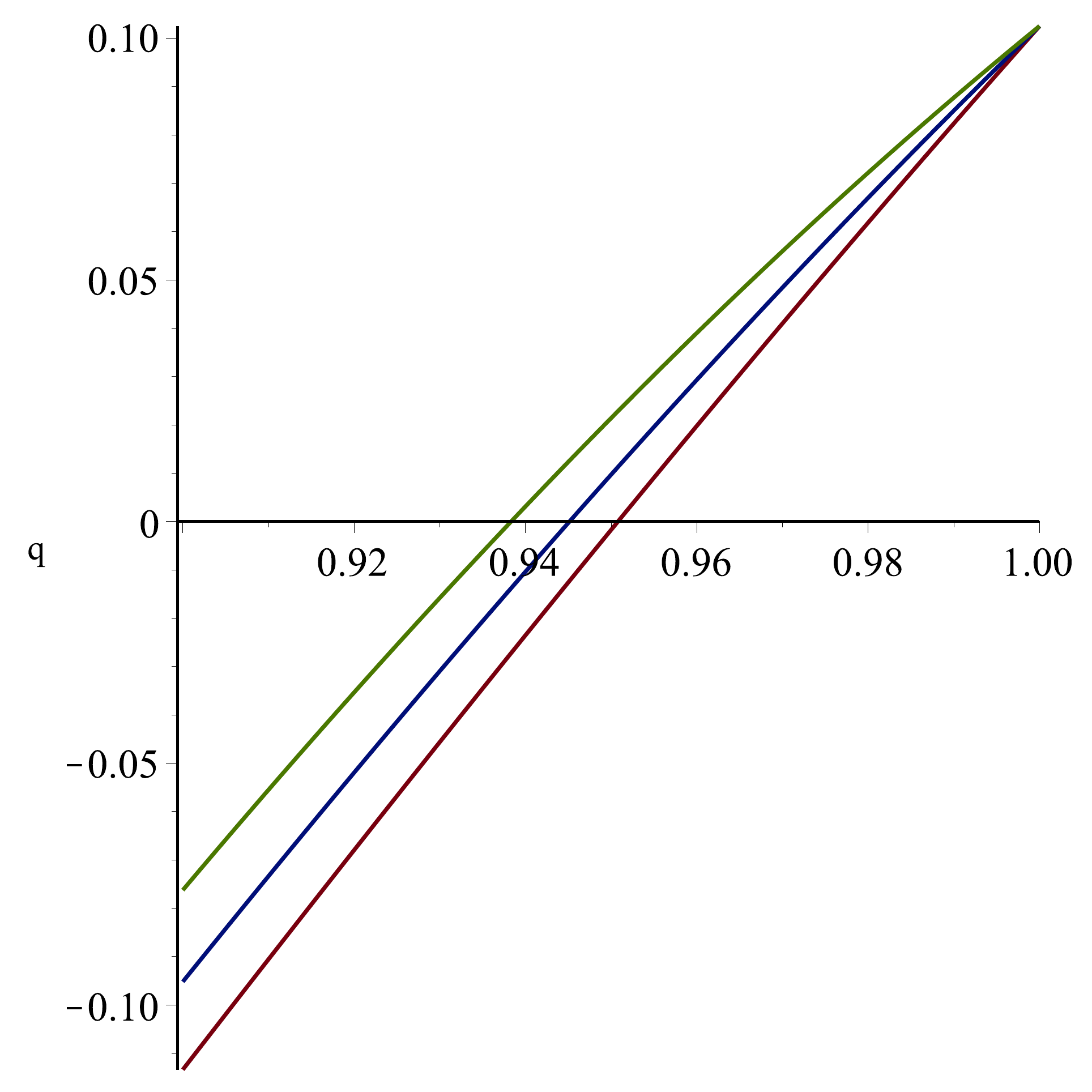}
        \caption{\small{The evolution of $q$ versus $\theta$ for various $n_2=1.5,~ 2.0,~2.5 $ (red, blue, green)}} \label{qn}
    \end{minipage}
\end{figure}

\begin{figure}[!htb]
    \begin{minipage}{0.48\textwidth}
        \centering
        \includegraphics[scale=0.42]{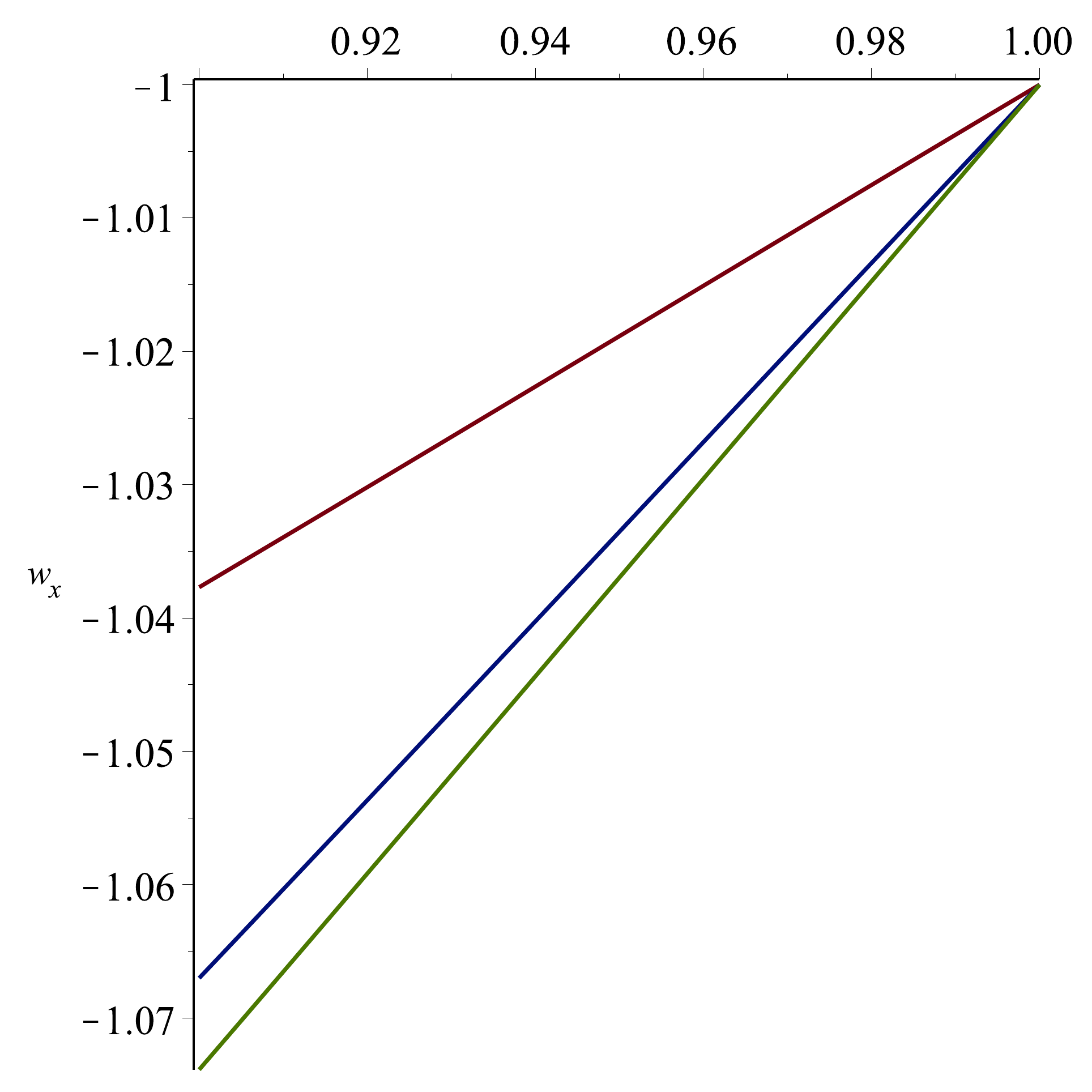}
        \caption{\small{The evolution of $w_x$ versus $\theta$ for various $b=0.5,~ 2/3,~ 0.7 $ (red, blue, green)}} \label{wb}
    \end{minipage}\hfill
     \begin{minipage}{0.48\textwidth}
        \centering
        \includegraphics[scale=0.42]{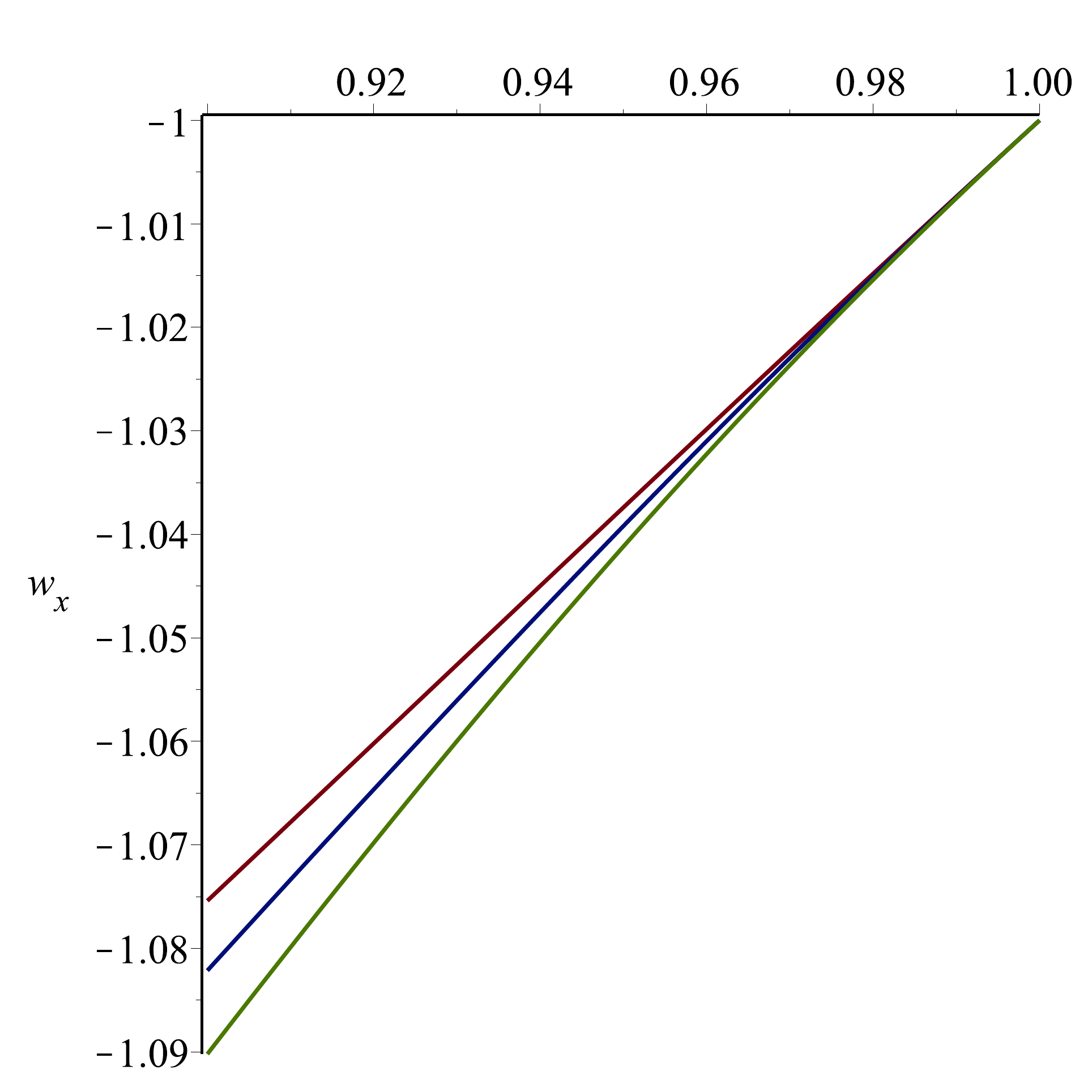}
        \caption{\small{The evolution of $w_x$ versus $\theta$ for various $n_2=2.0,~ 4.0,~6.0 $ (red, blue, green)}} \label{wn}
    \end{minipage}
\end{figure}

  Now substituting equations (\ref{eqint2}, \ref{HdH2}, \ref{rhod2}) in equation (\ref{eqint1}),
  the equation of state parameter $w_x$ as a function of $\theta$ around the turning point, $u \approx 0$, simply gives
 \bq
 w_x \approx-1-\frac{b^2}{\frac{2 n_2}{3 \Omega_x}+2\Omega_x(1-\epsilon)+4\epsilon-(\frac{n_2}{3}+3)}.\label{wx2}
 \eq
 and after a few simplifications it yields
 \bq
 w_x=-1-\frac{3 b^2 (1-\theta)}{(n_2-6\epsilon_1) (2 \theta^2-3\theta+1)-\theta+3 }.\label{wx3}
 \eq
 Also the deceleration parameter $q=1-\dot{H}/H^2$ can be calculated as
 \bq
 q=-1-\frac{9 b^2 (1-2\theta)}{2(n_2-6\epsilon_1)(2 \theta^2-3\theta+1)-2\theta+6}. \label{q3}
 \eq
Last equations show that in the non-interacting case, $b=0$, two
important functions $w_x=-1$ and $q=-1$ are constant and the turning
conditions were not retrieved. In fact, the shift from expansion to
contraction phases never happens. Therefore, the non-interacting
case should be abandoned altogether. Also, in order to turning from
expansion to contraction at a limit value, $\theta=1 $, Eq.
(\ref{q3}) requires a minimum value for the interaction term $ b=2/3
$. In all figures three functions $ H $, $w_x$ and $q$ are plotted
in $\theta$  (horizontal axis) around $ \theta=1 $. Fig \ref{He}
shows the effect of parameter $ \epsilon_1 $ in behavior of $ H $.
Increasing $ \epsilon_1 $ leads to an increase in the rate of $ H $
reduction. This parameter is related to parameter $ n_0 $ in DE
density. In Fig \ref{Hn}, the effect of parameter $ n_2 $ , second
term in DE density, in behavior of Hubble rate is plotted.
Increasing $ n_2 $ which reveals the effect of $ H^2 $ term in $
\rho_x $, leads to an increase in the rate of $ H $ reduction. Also,
for the concavity of the curve $ H $, with increasing $ n_2 $, it
changes from positive to negative. Fig \ref{qb} narrates the effect
of interaction parameter $ b $ in behavior of $ q $. This indicates
that for values $ b < 2/3 $, the value of $ q $ never vanished
(specially in non-interacting case $ b=0 $), and for values greater
than $ 2/3 $, turnaround time occurs earlier with increasing $ b $.
In Fig \ref{qn}, the effect of parameter $ n_2 $ , in behavior of $
q $ is shown. This indicates that an increase in $ n_2 $, reduces
 the transition time from
accelerating to decelerating expansion, and eventually in $
\theta=1 $, they all come together. And at last in Figs. \ref{wb},
\ref{wn}, the evolution of $ w_x $ are demonstrated. These show that
the universe behaves in phantom phase for all various of parameters
$ b $ and $ n_2 $, in a way that they merge to $w_x=-1 $ at $
\theta=1 $.

 It should be noted that because the crucial equations obtained in this analysis did not explicitly include n4 around the turnaround point, we could not show the weight effect of the term including H4 in the analysis. Of course, the effect of this sentence lies in the Epsilon1 coefficient.


% \begin{figure}[!]
%   \begin{center}
%       \includegraphics[scale=0.42]{figs/q.eps}
%       \caption{\small{The evolution of the conformal Hubble parameter  %$\mH$ versus $\eta$. At bounce point ($\eta=0;~~\mH =0$),
%               it shows a transition from a contracting universe to an %expanding one from left to right.}}
%       \label{H1}
%   \end{center}
% \end{figure}
\newpage
\section{concluding remarks}
After recent efforts, especially in 2021, on relic gravitational
waves from the merging of many black holes in the
collapsed phase of the universe in the pre-bounce regime, cyclic cosmology came to life again. Among models that are well-adapted to cosmological data, a deformed matter bounce scenario in which the matter is deformed by a component of RVM-DE is a good candidate for study. Our hypothesis is that just as at the time of bounce, where the energy density of radiation causes a critical density to reach the LQC energy level, in the turnaround phase, The dark energy that grows up to the critical energy, provides LQC energy condition. To prove this claim, we need a model of dark energy whose energy density can grow in the phantom phase to reach the critical energy and finally remains at the phantom wall in the rapid contraction phase of the universe. In this article, we seek to prove the capability of the RVM dark energy model in turnaround the universe from expanding to contracting. Following a previous study in which we found a successful case of the RVM-CDM model in which the running of the spectral index can accept a negative value, I was interested in one of the successful cases of the interacting model of the RVM-CDM, $\rho_x=n_0+n_2 H^2 +n_4 H^4$. The turnaround conditions at the end of expansion before the occurrence of a sudden big rip in a finite future were investigated.  At the turnaround point, the Hubble parameter must vanish, the EoS parameter must be equal to $ w_x = -1 $ for all choosing parameters, and the deceleration parameter must change from a negative to positive value near the turning point. After analyzing the Hubble parameter $H$, a new restriction between parameters of model was obtained. Also by considering the deceleration parameter and EoS parameter, around the turnaround, we obtained a minimum value for interaction parameter $ b=2/3 $. Therefore non-interacting models must be ignored  and can not create a successful turnaround. The universe behaves in the phantom regime, before any cyclicity and transition from an expansion to contraction at a finite time. At last the effect of each parameter in turnaround was studied and we saw that the transition time from accelerating to decelerating expansion occurs earlier for larger values of interaction parameter $ b $.
	Also in several graphs the effect of second term of RVM energy density, including $H^2$, was studied in parameter $n_2$.
	Increasing $n_2$, will reduce the time of transition point and will increase in the rate of $ H $ reduction. With this study, we showed that just as the RVM-CDM model can create a successful bounce, it can also be considered as a successful model for the future turnaround of the universe from an accelerated expansion to a contraction, in the presence of interaction.
\bibliography{khodamcyclic}

\begin{thebibliography}{73}
\expandafter\ifx\csname natexlab\endcsname\relax\def\natexlab#1{#1}\fi
\expandafter\ifx\csname bibnamefont\endcsname\relax
  \def\bibnamefont#1{#1}\fi
\expandafter\ifx\csname bibfnamefont\endcsname\relax
  \def\bibfnamefont#1{#1}\fi
\expandafter\ifx\csname citenamefont\endcsname\relax
  \def\citenamefont#1{#1}\fi
\expandafter\ifx\csname url\endcsname\relax
  \def\url#1{\texttt{#1}}\fi
\expandafter\ifx\csname urlprefix\endcsname\relax\def\urlprefix{URL }\fi
\providecommand{\bibinfo}[2]{#2}
\providecommand{\eprint}[2][]{\url{#2}}

\bibitem[{\citenamefont{Tolman}(1934)}]{Tolman_1934}
\bibinfo{author}{\bibfnamefont{R.~C.} \bibnamefont{Tolman}},
  \emph{\bibinfo{title}{{Relativity Thermodynamics and Cosmology}}}
  (\bibinfo{publisher}{Oxford, Clarendon Press}, \bibinfo{year}{1934}).

\bibitem[{\citenamefont{Caldwell et~al.}(2003)\citenamefont{Caldwell,
  Kamionkowski, and Weinberg}}]{Caldwell:2003vq}
\bibinfo{author}{\bibfnamefont{R.~R.} \bibnamefont{Caldwell}},
  \bibinfo{author}{\bibfnamefont{M.}~\bibnamefont{Kamionkowski}},
  \bibnamefont{and} \bibinfo{author}{\bibfnamefont{N.~N.}
  \bibnamefont{Weinberg}}, \bibinfo{journal}{Phys. Rev. Lett.}
  \textbf{\bibinfo{volume}{91}}, \bibinfo{pages}{071301}
  (\bibinfo{year}{2003}), \eprint{astro-ph/0302506}.

\bibitem[{\citenamefont{Nojiri et~al.}(2012)\citenamefont{Nojiri, Odintsov, and
  Saez-Gomez}}]{Nojiri:2011kd}
\bibinfo{author}{\bibfnamefont{S.}~\bibnamefont{Nojiri}},
  \bibinfo{author}{\bibfnamefont{S.~D.} \bibnamefont{Odintsov}},
  \bibnamefont{and}
  \bibinfo{author}{\bibfnamefont{D.}~\bibnamefont{Saez-Gomez}},
  \bibinfo{journal}{AIP Conf. Proc.} \textbf{\bibinfo{volume}{1458}},
  \bibinfo{pages}{207} (\bibinfo{year}{2012}), \eprint{1108.0767}.

\bibitem[{\citenamefont{Penrose}(2011)}]{Penrose2011CyclesOT}
\bibinfo{author}{\bibfnamefont{R.}~\bibnamefont{Penrose}},
  \emph{\bibinfo{title}{{Cycles of time: an extraordinary new view of the
  universe}}} (\bibinfo{publisher}{Bodley Head (UK), Knopf (US)},
  \bibinfo{year}{2011}).

\bibitem[{\citenamefont{Penrose}(2006)}]{Penrose:2006zz}
\bibinfo{author}{\bibfnamefont{R.}~\bibnamefont{Penrose}},
  \bibinfo{journal}{Conf. Proc. C} \textbf{\bibinfo{volume}{060626}},
  \bibinfo{pages}{2759} (\bibinfo{year}{2006}).

\bibitem[{\citenamefont{Penrose}(2012)}]{Penrose:2010zz}
\bibinfo{author}{\bibfnamefont{R.}~\bibnamefont{Penrose}},
  \bibinfo{journal}{AIP Conf. Proc.} \textbf{\bibinfo{volume}{1446}},
  \bibinfo{pages}{233} (\bibinfo{year}{2012}).

\bibitem[{\citenamefont{Gurzadyan and Penrose}(2011)}]{Gurzadyan:2011ac}
\bibinfo{author}{\bibfnamefont{V.~G.} \bibnamefont{Gurzadyan}}
  \bibnamefont{and} \bibinfo{author}{\bibfnamefont{R.}~\bibnamefont{Penrose}}
  (\bibinfo{year}{2011}), \eprint{1104.5675}.

\bibitem[{\citenamefont{Gurzadyan and Penrose}(2013)}]{Gurzadyan:2013cna}
\bibinfo{author}{\bibfnamefont{V.~G.} \bibnamefont{Gurzadyan}}
  \bibnamefont{and} \bibinfo{author}{\bibfnamefont{R.}~\bibnamefont{Penrose}},
  \bibinfo{journal}{Eur. Phys. J. Plus} \textbf{\bibinfo{volume}{128}},
  \bibinfo{pages}{22} (\bibinfo{year}{2013}), \eprint{1302.5162}.

\bibitem[{\citenamefont{Gurzadyan and Penrose}(2016)}]{Gurzadyan:2015gzb}
\bibinfo{author}{\bibfnamefont{V.~G.} \bibnamefont{Gurzadyan}}
  \bibnamefont{and} \bibinfo{author}{\bibfnamefont{R.}~\bibnamefont{Penrose}},
  \bibinfo{journal}{Eur. Phys. J. Plus} \textbf{\bibinfo{volume}{131}},
  \bibinfo{pages}{11} (\bibinfo{year}{2016}), \eprint{1512.00554}.

\bibitem[{\citenamefont{Penrose}(2017)}]{Penrose:2017npq}
\bibinfo{author}{\bibfnamefont{R.}~\bibnamefont{Penrose}}
  (\bibinfo{year}{2017}), \eprint{1707.04169}.

\bibitem[{\citenamefont{An et~al.}(2020)\citenamefont{An, Meissner, Nurowski,
  and Penrose}}]{An:2018utx}
\bibinfo{author}{\bibfnamefont{D.}~\bibnamefont{An}},
  \bibinfo{author}{\bibfnamefont{K.~A.} \bibnamefont{Meissner}},
  \bibinfo{author}{\bibfnamefont{P.}~\bibnamefont{Nurowski}}, \bibnamefont{and}
  \bibinfo{author}{\bibfnamefont{R.}~\bibnamefont{Penrose}},
  \bibinfo{journal}{Mon. Not. Roy. Astron. Soc.}
  \textbf{\bibinfo{volume}{495}}, \bibinfo{pages}{3403} (\bibinfo{year}{2020}),
  \eprint{1808.01740}.

\bibitem[{\citenamefont{Steinhardt and Turok}(2002)}]{Steinhardt:2001st}
\bibinfo{author}{\bibfnamefont{P.~J.} \bibnamefont{Steinhardt}}
  \bibnamefont{and} \bibinfo{author}{\bibfnamefont{N.}~\bibnamefont{Turok}},
  \bibinfo{journal}{Phys. Rev. D} \textbf{\bibinfo{volume}{65}},
  \bibinfo{pages}{126003} (\bibinfo{year}{2002}), \eprint{hep-th/0111098}.

\bibitem[{\citenamefont{Steinhardt et~al.}(2002)\citenamefont{Steinhardt,
  Turok, and Turok}}]{Steinhardt:2002ih}
\bibinfo{author}{\bibfnamefont{P.~J.} \bibnamefont{Steinhardt}},
  \bibinfo{author}{\bibfnamefont{N.}~\bibnamefont{Turok}}, \bibnamefont{and}
  \bibinfo{author}{\bibfnamefont{N.}~\bibnamefont{Turok}},
  \bibinfo{journal}{Science} \textbf{\bibinfo{volume}{296}},
  \bibinfo{pages}{1436} (\bibinfo{year}{2002}), \eprint{hep-th/0111030}.

\bibitem[{\citenamefont{Steinhardt and Turok}(2003)}]{Steinhardt:2002kw}
\bibinfo{author}{\bibfnamefont{P.~J.} \bibnamefont{Steinhardt}}
  \bibnamefont{and} \bibinfo{author}{\bibfnamefont{N.}~\bibnamefont{Turok}},
  \bibinfo{journal}{Nucl. Phys. B Proc. Suppl.} \textbf{\bibinfo{volume}{124}},
  \bibinfo{pages}{38} (\bibinfo{year}{2003}), \eprint{astro-ph/0204479}.

\bibitem[{\citenamefont{Boyle et~al.}(2004)\citenamefont{Boyle, Steinhardt, and
  Turok}}]{Boyle:2003km}
\bibinfo{author}{\bibfnamefont{L.~A.} \bibnamefont{Boyle}},
  \bibinfo{author}{\bibfnamefont{P.~J.} \bibnamefont{Steinhardt}},
  \bibnamefont{and} \bibinfo{author}{\bibfnamefont{N.}~\bibnamefont{Turok}},
  \bibinfo{journal}{Phys. Rev. D} \textbf{\bibinfo{volume}{69}},
  \bibinfo{pages}{127302} (\bibinfo{year}{2004}), \eprint{hep-th/0307170}.

\bibitem[{\citenamefont{Khoury et~al.}(2004)\citenamefont{Khoury, Steinhardt,
  and Turok}}]{Khoury:2003rt}
\bibinfo{author}{\bibfnamefont{J.}~\bibnamefont{Khoury}},
  \bibinfo{author}{\bibfnamefont{P.~J.} \bibnamefont{Steinhardt}},
  \bibnamefont{and} \bibinfo{author}{\bibfnamefont{N.}~\bibnamefont{Turok}},
  \bibinfo{journal}{Phys. Rev. Lett.} \textbf{\bibinfo{volume}{92}},
  \bibinfo{pages}{031302} (\bibinfo{year}{2004}), \eprint{hep-th/0307132}.

\bibitem[{\citenamefont{Steinhardt and Turok}(2005)}]{Steinhardt:2004gk}
\bibinfo{author}{\bibfnamefont{P.~J.} \bibnamefont{Steinhardt}}
  \bibnamefont{and} \bibinfo{author}{\bibfnamefont{N.}~\bibnamefont{Turok}},
  \bibinfo{journal}{New Astron. Rev.} \textbf{\bibinfo{volume}{49}},
  \bibinfo{pages}{43} (\bibinfo{year}{2005}), \eprint{astro-ph/0404480}.

\bibitem[{\citenamefont{Lehners and Steinhardt}(2008)}]{Lehners:2008my}
\bibinfo{author}{\bibfnamefont{J.-L.} \bibnamefont{Lehners}} \bibnamefont{and}
  \bibinfo{author}{\bibfnamefont{P.~J.} \bibnamefont{Steinhardt}},
  \bibinfo{journal}{Phys. Rev. D} \textbf{\bibinfo{volume}{78}},
  \bibinfo{pages}{023506} (\bibinfo{year}{2008}), \bibinfo{note}{[Erratum:
  Phys.Rev.D 79, 129902 (2009)]}, \eprint{0804.1293}.

\bibitem[{\citenamefont{Lehners and Steinhardt}(2013)}]{Lehners:2013cka}
\bibinfo{author}{\bibfnamefont{J.-L.} \bibnamefont{Lehners}} \bibnamefont{and}
  \bibinfo{author}{\bibfnamefont{P.~J.} \bibnamefont{Steinhardt}},
  \bibinfo{journal}{Phys. Rev. D} \textbf{\bibinfo{volume}{87}},
  \bibinfo{pages}{123533} (\bibinfo{year}{2013}), \eprint{1304.3122}.

\bibitem[{\citenamefont{Ijjas and Steinhardt}(2016)}]{Ijjas:2016tpn}
\bibinfo{author}{\bibfnamefont{A.}~\bibnamefont{Ijjas}} \bibnamefont{and}
  \bibinfo{author}{\bibfnamefont{P.~J.} \bibnamefont{Steinhardt}},
  \bibinfo{journal}{Phys. Rev. Lett.} \textbf{\bibinfo{volume}{117}},
  \bibinfo{pages}{121304} (\bibinfo{year}{2016}), \eprint{1606.08880}.

\bibitem[{\citenamefont{Ijjas and Steinhardt}(2017)}]{Ijjas:2016vtq}
\bibinfo{author}{\bibfnamefont{A.}~\bibnamefont{Ijjas}} \bibnamefont{and}
  \bibinfo{author}{\bibfnamefont{P.~J.} \bibnamefont{Steinhardt}},
  \bibinfo{journal}{Phys. Lett. B} \textbf{\bibinfo{volume}{764}},
  \bibinfo{pages}{289} (\bibinfo{year}{2017}), \eprint{1609.01253}.

\bibitem[{\citenamefont{Ijjas and Steinhardt}(2018)}]{Ijjas:2018qbo}
\bibinfo{author}{\bibfnamefont{A.}~\bibnamefont{Ijjas}} \bibnamefont{and}
  \bibinfo{author}{\bibfnamefont{P.~J.} \bibnamefont{Steinhardt}},
  \bibinfo{journal}{Class. Quant. Grav.} \textbf{\bibinfo{volume}{35}},
  \bibinfo{pages}{135004} (\bibinfo{year}{2018}), \eprint{1803.01961}.

\bibitem[{\citenamefont{Ijjas and Steinhardt}(2022)}]{Ijjas:2021zwv}
\bibinfo{author}{\bibfnamefont{A.}~\bibnamefont{Ijjas}} \bibnamefont{and}
  \bibinfo{author}{\bibfnamefont{P.~J.} \bibnamefont{Steinhardt}},
  \bibinfo{journal}{Phys. Lett. B} \textbf{\bibinfo{volume}{824}},
  \bibinfo{pages}{136823} (\bibinfo{year}{2022}), \eprint{2108.07101}.

\bibitem[{\citenamefont{Ijjas and Steinhardt}(2019)}]{Ijjas:2019pyf}
\bibinfo{author}{\bibfnamefont{A.}~\bibnamefont{Ijjas}} \bibnamefont{and}
  \bibinfo{author}{\bibfnamefont{P.~J.} \bibnamefont{Steinhardt}},
  \bibinfo{journal}{Phys. Lett. B} \textbf{\bibinfo{volume}{795}},
  \bibinfo{pages}{666} (\bibinfo{year}{2019}), \eprint{1904.08022}.

\bibitem[{\citenamefont{Baum and Frampton}(2007)}]{Baum:2006ee}
\bibinfo{author}{\bibfnamefont{L.}~\bibnamefont{Baum}} \bibnamefont{and}
  \bibinfo{author}{\bibfnamefont{P.~H.} \bibnamefont{Frampton}},
  \bibinfo{journal}{Phys. Rev. Lett.} \textbf{\bibinfo{volume}{98}},
  \bibinfo{pages}{071301} (\bibinfo{year}{2007}), \eprint{hep-th/0610213}.

\bibitem[{\citenamefont{Frampton}(2006)}]{Frampton:2006mg}
\bibinfo{author}{\bibfnamefont{P.~H.} \bibnamefont{Frampton}}, in
  \emph{\bibinfo{booktitle}{{2006 International Workshop on the Origin of Mass
  and Strong Coupling Gauge Theories (SCGT 06)}}} (\bibinfo{year}{2006}), pp.
  \bibinfo{pages}{331--337}, \eprint{astro-ph/0612243}.

\bibitem[{\citenamefont{Frampton}(2007)}]{Frampton:2007cv}
\bibinfo{author}{\bibfnamefont{P.~H.} \bibnamefont{Frampton}},
  \bibinfo{journal}{Mod. Phys. Lett. A} \textbf{\bibinfo{volume}{22}},
  \bibinfo{pages}{2587} (\bibinfo{year}{2007}), \eprint{0705.2730}.

\bibitem[{\citenamefont{Aref'eva et~al.}(2009)\citenamefont{Aref'eva, Frampton,
  and Matsuzaki}}]{Arefeva:2008abt}
\bibinfo{author}{\bibfnamefont{I.}~\bibnamefont{Aref'eva}},
  \bibinfo{author}{\bibfnamefont{P.~H.} \bibnamefont{Frampton}},
  \bibnamefont{and}
  \bibinfo{author}{\bibfnamefont{S.}~\bibnamefont{Matsuzaki}},
  \bibinfo{journal}{Proc. Steklov Inst. Math.} \textbf{\bibinfo{volume}{265}},
  \bibinfo{pages}{59} (\bibinfo{year}{2009}), \eprint{0802.1294}.

\bibitem[{\citenamefont{Frampton and Ludwick}(2013)}]{Frampton:2013fxa}
\bibinfo{author}{\bibfnamefont{P.~H.} \bibnamefont{Frampton}} \bibnamefont{and}
  \bibinfo{author}{\bibfnamefont{K.~J.} \bibnamefont{Ludwick}},
  \bibinfo{journal}{Mod. Phys. Lett. A} \textbf{\bibinfo{volume}{28}},
  \bibinfo{pages}{1350125} (\bibinfo{year}{2013}), \eprint{1304.5221}.

\bibitem[{\citenamefont{Corian\`o and Frampton}(2019)}]{Coriano:2019eif}
\bibinfo{author}{\bibfnamefont{C.}~\bibnamefont{Corian\`o}} \bibnamefont{and}
  \bibinfo{author}{\bibfnamefont{P.~H.} \bibnamefont{Frampton}},
  \bibinfo{journal}{Mod. Phys. Lett. A} \textbf{\bibinfo{volume}{35}},
  \bibinfo{pages}{1950355} (\bibinfo{year}{2019}), \eprint{1906.10090}.

\bibitem[{\citenamefont{Singh}(2009)}]{Singh:2009mz}
\bibinfo{author}{\bibfnamefont{P.}~\bibnamefont{Singh}},
  \bibinfo{journal}{Class. Quant. Grav.} \textbf{\bibinfo{volume}{26}},
  \bibinfo{pages}{125005} (\bibinfo{year}{2009}), \eprint{0901.2750}.

\bibitem[{\citenamefont{Ashtekar and Singh}(2011)}]{Ashtekar:2011ni}
\bibinfo{author}{\bibfnamefont{A.}~\bibnamefont{Ashtekar}} \bibnamefont{and}
  \bibinfo{author}{\bibfnamefont{P.}~\bibnamefont{Singh}},
  \bibinfo{journal}{Class. Quant. Grav.} \textbf{\bibinfo{volume}{28}},
  \bibinfo{pages}{213001} (\bibinfo{year}{2011}), \eprint{1108.0893}.

\bibitem[{\citenamefont{Wilson-Ewing}(2013)}]{WilsonEwing:2012pu}
\bibinfo{author}{\bibfnamefont{E.}~\bibnamefont{Wilson-Ewing}},
  \bibinfo{journal}{JCAP} \textbf{\bibinfo{volume}{1303}}, \bibinfo{pages}{026}
  (\bibinfo{year}{2013}), \eprint{1211.6269}.

\bibitem[{\citenamefont{Gorkavyi}(2022)}]{Gorkavyi:2021tbw}
\bibinfo{author}{\bibfnamefont{N.}~\bibnamefont{Gorkavyi}},
  \bibinfo{journal}{New Astron.} \textbf{\bibinfo{volume}{91}},
  \bibinfo{pages}{101698} (\bibinfo{year}{2022}), \eprint{2110.10218}.

\bibitem[{\citenamefont{Gasperini and Veneziano}(1993)}]{Gasperini:1992em}
\bibinfo{author}{\bibfnamefont{M.}~\bibnamefont{Gasperini}} \bibnamefont{and}
  \bibinfo{author}{\bibfnamefont{G.}~\bibnamefont{Veneziano}},
  \bibinfo{journal}{Astropart. Phys.} \textbf{\bibinfo{volume}{1}},
  \bibinfo{pages}{317} (\bibinfo{year}{1993}), \eprint{hep-th/9211021}.

\bibitem[{\citenamefont{Khoury et~al.}(2001)\citenamefont{Khoury, Ovrut,
  Steinhardt, and Turok}}]{Khoury:2001wf}
\bibinfo{author}{\bibfnamefont{J.}~\bibnamefont{Khoury}},
  \bibinfo{author}{\bibfnamefont{B.~A.} \bibnamefont{Ovrut}},
  \bibinfo{author}{\bibfnamefont{P.~J.} \bibnamefont{Steinhardt}},
  \bibnamefont{and} \bibinfo{author}{\bibfnamefont{N.}~\bibnamefont{Turok}},
  \bibinfo{journal}{Phys. Rev.} \textbf{\bibinfo{volume}{D64}},
  \bibinfo{pages}{123522} (\bibinfo{year}{2001}), \eprint{hep-th/0103239}.

\bibitem[{\citenamefont{Brandenberger and Vafa}(1989)}]{Brandenberger:1988aj}
\bibinfo{author}{\bibfnamefont{R.~H.} \bibnamefont{Brandenberger}}
  \bibnamefont{and} \bibinfo{author}{\bibfnamefont{C.}~\bibnamefont{Vafa}},
  \bibinfo{journal}{Nucl. Phys.} \textbf{\bibinfo{volume}{B316}},
  \bibinfo{pages}{391} (\bibinfo{year}{1989}).

\bibitem[{\citenamefont{Nayeri et~al.}(2006)\citenamefont{Nayeri,
  Brandenberger, and Vafa}}]{Nayeri:2005ck}
\bibinfo{author}{\bibfnamefont{A.}~\bibnamefont{Nayeri}},
  \bibinfo{author}{\bibfnamefont{R.~H.} \bibnamefont{Brandenberger}},
  \bibnamefont{and} \bibinfo{author}{\bibfnamefont{C.}~\bibnamefont{Vafa}},
  \bibinfo{journal}{Phys. Rev. Lett.} \textbf{\bibinfo{volume}{97}},
  \bibinfo{pages}{021302} (\bibinfo{year}{2006}), \eprint{hep-th/0511140}.

\bibitem[{\citenamefont{Finelli and Brandenberger}(2002)}]{Finelli:2001sr}
\bibinfo{author}{\bibfnamefont{F.}~\bibnamefont{Finelli}} \bibnamefont{and}
  \bibinfo{author}{\bibfnamefont{R.}~\bibnamefont{Brandenberger}},
  \bibinfo{journal}{Phys. Rev.} \textbf{\bibinfo{volume}{D65}},
  \bibinfo{pages}{103522} (\bibinfo{year}{2002}), \eprint{hep-th/0112249}.

\bibitem[{\citenamefont{Lin et~al.}(2011)\citenamefont{Lin, Brandenberger, and
  Perreault~Levasseur}}]{Lin:2010pf}
\bibinfo{author}{\bibfnamefont{C.}~\bibnamefont{Lin}},
  \bibinfo{author}{\bibfnamefont{R.~H.} \bibnamefont{Brandenberger}},
  \bibnamefont{and}
  \bibinfo{author}{\bibfnamefont{L.}~\bibnamefont{Perreault~Levasseur}},
  \bibinfo{journal}{JCAP} \textbf{\bibinfo{volume}{1104}}, \bibinfo{pages}{019}
  (\bibinfo{year}{2011}), \eprint{1007.2654}.

\bibitem[{\citenamefont{Cai et~al.}(2013)\citenamefont{Cai, McDonough,
  Duplessis, and Brandenberger}}]{Cai:2013kja}
\bibinfo{author}{\bibfnamefont{Y.-F.} \bibnamefont{Cai}},
  \bibinfo{author}{\bibfnamefont{E.}~\bibnamefont{McDonough}},
  \bibinfo{author}{\bibfnamefont{F.}~\bibnamefont{Duplessis}},
  \bibnamefont{and} \bibinfo{author}{\bibfnamefont{R.~H.}
  \bibnamefont{Brandenberger}}, \bibinfo{journal}{JCAP}
  \textbf{\bibinfo{volume}{1310}}, \bibinfo{pages}{024} (\bibinfo{year}{2013}),
  \eprint{1305.5259}.

\bibitem[{\citenamefont{Lehners and Wilson-Ewing}(2015)}]{Lehners:2015mra}
\bibinfo{author}{\bibfnamefont{J.-L.} \bibnamefont{Lehners}} \bibnamefont{and}
  \bibinfo{author}{\bibfnamefont{E.}~\bibnamefont{Wilson-Ewing}},
  \bibinfo{journal}{JCAP} \textbf{\bibinfo{volume}{10}}, \bibinfo{pages}{038}
  (\bibinfo{year}{2015}), \eprint{1507.08112}.

\bibitem[{\citenamefont{Cai et~al.}(2016{\natexlab{a}})\citenamefont{Cai,
  Duplessis, Easson, and Wang}}]{Cai:2015vzv}
\bibinfo{author}{\bibfnamefont{Y.-F.} \bibnamefont{Cai}},
  \bibinfo{author}{\bibfnamefont{F.}~\bibnamefont{Duplessis}},
  \bibinfo{author}{\bibfnamefont{D.~A.} \bibnamefont{Easson}},
  \bibnamefont{and} \bibinfo{author}{\bibfnamefont{D.-G.} \bibnamefont{Wang}},
  \bibinfo{journal}{Phys. Rev.} \textbf{\bibinfo{volume}{D93}},
  \bibinfo{pages}{043546} (\bibinfo{year}{2016}{\natexlab{a}}),
  \eprint{1512.08979}.

\bibitem[{\citenamefont{Cai et~al.}(2016{\natexlab{b}})\citenamefont{Cai,
  Marciano, Wang, and Wilson-Ewing}}]{Cai:2016hea}
\bibinfo{author}{\bibfnamefont{Y.-F.} \bibnamefont{Cai}},
  \bibinfo{author}{\bibfnamefont{A.}~\bibnamefont{Marciano}},
  \bibinfo{author}{\bibfnamefont{D.-G.} \bibnamefont{Wang}}, \bibnamefont{and}
  \bibinfo{author}{\bibfnamefont{E.}~\bibnamefont{Wilson-Ewing}},
  \bibinfo{journal}{Universe} \textbf{\bibinfo{volume}{3}}, \bibinfo{pages}{1}
  (\bibinfo{year}{2016}{\natexlab{b}}), \eprint{1610.00938}.

\bibitem[{\citenamefont{Ade et~al.}(2016{\natexlab{a}})}]{Planck:2015sxf}
\bibinfo{author}{\bibfnamefont{P.~A.~R.} \bibnamefont{Ade}}
  \bibnamefont{et~al.} (\bibinfo{collaboration}{Planck}),
  \bibinfo{journal}{Astron. Astrophys.} \textbf{\bibinfo{volume}{594}},
  \bibinfo{pages}{A20} (\bibinfo{year}{2016}{\natexlab{a}}),
  \eprint{1502.02114}.

\bibitem[{\citenamefont{Ade et~al.}(2016{\natexlab{b}})}]{Planck:2015fie}
\bibinfo{author}{\bibfnamefont{P.~A.~R.} \bibnamefont{Ade}}
  \bibnamefont{et~al.} (\bibinfo{collaboration}{Planck}),
  \bibinfo{journal}{Astron. Astrophys.} \textbf{\bibinfo{volume}{594}},
  \bibinfo{pages}{A13} (\bibinfo{year}{2016}{\natexlab{b}}),
  \eprint{1502.01589}.

\bibitem[{\citenamefont{Ade et~al.}(2015)}]{Ade:2015tva}
\bibinfo{author}{\bibfnamefont{P.~A.~R.} \bibnamefont{Ade}}
  \bibnamefont{et~al.} (\bibinfo{collaboration}{BICEP2, Planck}),
  \bibinfo{journal}{Phys. Rev. Lett.} \textbf{\bibinfo{volume}{114}},
  \bibinfo{pages}{101301} (\bibinfo{year}{2015}), \eprint{1502.00612}.

\bibitem[{\citenamefont{Ade et~al.}(2018)}]{BICEP2:2018kqh}
\bibinfo{author}{\bibfnamefont{P.~A.~R.} \bibnamefont{Ade}}
  \bibnamefont{et~al.} (\bibinfo{collaboration}{BICEP2, Keck Array}),
  \bibinfo{journal}{Phys. Rev. Lett.} \textbf{\bibinfo{volume}{121}},
  \bibinfo{pages}{221301} (\bibinfo{year}{2018}), \eprint{1810.05216}.

\bibitem[{\citenamefont{Cai and Wilson-Ewing}(2014)}]{Cai:2014zga}
\bibinfo{author}{\bibfnamefont{Y.-F.} \bibnamefont{Cai}} \bibnamefont{and}
  \bibinfo{author}{\bibfnamefont{E.}~\bibnamefont{Wilson-Ewing}},
  \bibinfo{journal}{JCAP} \textbf{\bibinfo{volume}{1403}}, \bibinfo{pages}{026}
  (\bibinfo{year}{2014}), \eprint{1402.3009}.

\bibitem[{\citenamefont{Cai and Wilson-Ewing}(2015)}]{Cai:2014jla}
\bibinfo{author}{\bibfnamefont{Y.-F.} \bibnamefont{Cai}} \bibnamefont{and}
  \bibinfo{author}{\bibfnamefont{E.}~\bibnamefont{Wilson-Ewing}},
  \bibinfo{journal}{JCAP} \textbf{\bibinfo{volume}{1503}}, \bibinfo{pages}{006}
  (\bibinfo{year}{2015}), \eprint{1412.2914}.

\bibitem[{\citenamefont{Arab and Khodam-Mohammadi}(2018)}]{Arab:2017gae}
\bibinfo{author}{\bibfnamefont{M.}~\bibnamefont{Arab}} \bibnamefont{and}
  \bibinfo{author}{\bibfnamefont{A.}~\bibnamefont{Khodam-Mohammadi}},
  \bibinfo{journal}{Eur. Phys. J. C} \textbf{\bibinfo{volume}{78}},
  \bibinfo{pages}{243} (\bibinfo{year}{2018}), \eprint{1707.06464}.

\bibitem[{\citenamefont{Arab and Khorasani}(2021)}]{Arab:2021wis}
\bibinfo{author}{\bibfnamefont{M.}~\bibnamefont{Arab}} \bibnamefont{and}
  \bibinfo{author}{\bibfnamefont{M.}~\bibnamefont{Khorasani}}
  (\bibinfo{year}{2021}), \eprint{2107.08331}.

\bibitem[{\citenamefont{Zhang et~al.}(2007)\citenamefont{Zhang, Zhang, and
  Liu}}]{Zhang:2007an}
\bibinfo{author}{\bibfnamefont{J.-f.} \bibnamefont{Zhang}},
  \bibinfo{author}{\bibfnamefont{X.}~\bibnamefont{Zhang}}, \bibnamefont{and}
  \bibinfo{author}{\bibfnamefont{H.-y.} \bibnamefont{Liu}},
  \bibinfo{journal}{Eur. Phys. J. C} \textbf{\bibinfo{volume}{52}},
  \bibinfo{pages}{693} (\bibinfo{year}{2007}), \eprint{0708.3121}.

\bibitem[{\citenamefont{Sheykhi et~al.}(2018)\citenamefont{Sheykhi, Tavayef,
  and Moradpour}}]{Sheykhi:2017cid}
\bibinfo{author}{\bibfnamefont{A.}~\bibnamefont{Sheykhi}},
  \bibinfo{author}{\bibfnamefont{M.}~\bibnamefont{Tavayef}}, \bibnamefont{and}
  \bibinfo{author}{\bibfnamefont{H.}~\bibnamefont{Moradpour}},
  \bibinfo{journal}{Can. J. Phys.} \textbf{\bibinfo{volume}{96}},
  \bibinfo{pages}{1034} (\bibinfo{year}{2018}), \eprint{1706.04433}.

\bibitem[{\citenamefont{Ashtekar et~al.}(2006)\citenamefont{Ashtekar,
  Pawlowski, and Singh}}]{Ashtekar:2006wn}
\bibinfo{author}{\bibfnamefont{A.}~\bibnamefont{Ashtekar}},
  \bibinfo{author}{\bibfnamefont{T.}~\bibnamefont{Pawlowski}},
  \bibnamefont{and} \bibinfo{author}{\bibfnamefont{P.}~\bibnamefont{Singh}},
  \bibinfo{journal}{Phys. Rev.} \textbf{\bibinfo{volume}{D74}},
  \bibinfo{pages}{084003} (\bibinfo{year}{2006}), \eprint{gr-qc/0607039}.

\bibitem[{\citenamefont{Perico et~al.}(2013)\citenamefont{Perico, Lima,
  Basilakos, and Sola}}]{Perico:2013mna}
\bibinfo{author}{\bibfnamefont{E.~L.~D.} \bibnamefont{Perico}},
  \bibinfo{author}{\bibfnamefont{J.~A.~S.} \bibnamefont{Lima}},
  \bibinfo{author}{\bibfnamefont{S.}~\bibnamefont{Basilakos}},
  \bibnamefont{and} \bibinfo{author}{\bibfnamefont{J.}~\bibnamefont{Sola}},
  \bibinfo{journal}{Phys. Rev. D} \textbf{\bibinfo{volume}{88}},
  \bibinfo{pages}{063531} (\bibinfo{year}{2013}), \eprint{1306.0591}.

\bibitem[{\citenamefont{Shapiro and Sola}(2000)}]{Shapiro:1999zt}
\bibinfo{author}{\bibfnamefont{I.~L.} \bibnamefont{Shapiro}} \bibnamefont{and}
  \bibinfo{author}{\bibfnamefont{J.}~\bibnamefont{Sola}},
  \bibinfo{journal}{Phys. Lett. B} \textbf{\bibinfo{volume}{475}},
  \bibinfo{pages}{236} (\bibinfo{year}{2000}), \eprint{hep-ph/9910462}.

\bibitem[{\citenamefont{Shapiro and Sola}(2002)}]{Shapiro:2000dz}
\bibinfo{author}{\bibfnamefont{I.~L.} \bibnamefont{Shapiro}} \bibnamefont{and}
  \bibinfo{author}{\bibfnamefont{J.}~\bibnamefont{Sola}},
  \bibinfo{journal}{JHEP} \textbf{\bibinfo{volume}{02}}, \bibinfo{pages}{006}
  (\bibinfo{year}{2002}), \eprint{hep-th/0012227}.

\bibitem[{\citenamefont{Babic et~al.}(2005)\citenamefont{Babic, Guberina,
  Horvat, and Stefancic}}]{Babic:2004ev}
\bibinfo{author}{\bibfnamefont{A.}~\bibnamefont{Babic}},
  \bibinfo{author}{\bibfnamefont{B.}~\bibnamefont{Guberina}},
  \bibinfo{author}{\bibfnamefont{R.}~\bibnamefont{Horvat}}, \bibnamefont{and}
  \bibinfo{author}{\bibfnamefont{H.}~\bibnamefont{Stefancic}},
  \bibinfo{journal}{Phys. Rev. D} \textbf{\bibinfo{volume}{71}},
  \bibinfo{pages}{124041} (\bibinfo{year}{2005}), \eprint{astro-ph/0407572}.

\bibitem[{\citenamefont{Shapiro et~al.}(2003)\citenamefont{Shapiro, Sola,
  Espana-Bonet, and Ruiz-Lapuente}}]{Shapiro:2003ui}
\bibinfo{author}{\bibfnamefont{I.~L.} \bibnamefont{Shapiro}},
  \bibinfo{author}{\bibfnamefont{J.}~\bibnamefont{Sola}},
  \bibinfo{author}{\bibfnamefont{C.}~\bibnamefont{Espana-Bonet}},
  \bibnamefont{and}
  \bibinfo{author}{\bibfnamefont{P.}~\bibnamefont{Ruiz-Lapuente}},
  \bibinfo{journal}{Phys. Lett. B} \textbf{\bibinfo{volume}{574}},
  \bibinfo{pages}{149} (\bibinfo{year}{2003}), \eprint{astro-ph/0303306}.

\bibitem[{\citenamefont{Espana-Bonet et~al.}(2004)\citenamefont{Espana-Bonet,
  Ruiz-Lapuente, Shapiro, and Sola}}]{Espana-Bonet:2003qjh}
\bibinfo{author}{\bibfnamefont{C.}~\bibnamefont{Espana-Bonet}},
  \bibinfo{author}{\bibfnamefont{P.}~\bibnamefont{Ruiz-Lapuente}},
  \bibinfo{author}{\bibfnamefont{I.~L.} \bibnamefont{Shapiro}},
  \bibnamefont{and} \bibinfo{author}{\bibfnamefont{J.}~\bibnamefont{Sola}},
  \bibinfo{journal}{JCAP} \textbf{\bibinfo{volume}{02}}, \bibinfo{pages}{006}
  (\bibinfo{year}{2004}), \eprint{hep-ph/0311171}.

\bibitem[{\citenamefont{Shapiro and Sola}(2004)}]{Shapiro:2003kv}
\bibinfo{author}{\bibfnamefont{I.~L.} \bibnamefont{Shapiro}} \bibnamefont{and}
  \bibinfo{author}{\bibfnamefont{J.}~\bibnamefont{Sola}},
  \bibinfo{journal}{Nucl. Phys. B Proc. Suppl.} \textbf{\bibinfo{volume}{127}},
  \bibinfo{pages}{71} (\bibinfo{year}{2004}), \eprint{hep-ph/0305279}.

\bibitem[{\citenamefont{Sola}(2008)}]{Sola:2007sv}
\bibinfo{author}{\bibfnamefont{J.}~\bibnamefont{Sola}}, \bibinfo{journal}{J.
  Phys. A} \textbf{\bibinfo{volume}{41}}, \bibinfo{pages}{164066}
  (\bibinfo{year}{2008}), \eprint{0710.4151}.

\bibitem[{\citenamefont{Shapiro and Sola}(2009)}]{Shapiro:2009dh}
\bibinfo{author}{\bibfnamefont{I.~L.} \bibnamefont{Shapiro}} \bibnamefont{and}
  \bibinfo{author}{\bibfnamefont{J.}~\bibnamefont{Sola}},
  \bibinfo{journal}{Phys. Lett. B} \textbf{\bibinfo{volume}{682}},
  \bibinfo{pages}{105} (\bibinfo{year}{2009}), \eprint{0910.4925}.

\bibitem[{\citenamefont{Basilakos et~al.}(2009)\citenamefont{Basilakos,
  Plionis, and Sol\`a}}]{Basilakos:2009wi}
\bibinfo{author}{\bibfnamefont{S.}~\bibnamefont{Basilakos}},
  \bibinfo{author}{\bibfnamefont{M.}~\bibnamefont{Plionis}}, \bibnamefont{and}
  \bibinfo{author}{\bibfnamefont{J.}~\bibnamefont{Sol\`a}},
  \bibinfo{journal}{Phys. Rev. D} \textbf{\bibinfo{volume}{80}},
  \bibinfo{pages}{083511} (\bibinfo{year}{2009}), \eprint{0907.4555}.

\bibitem[{\citenamefont{Grande et~al.}(2011)\citenamefont{Grande, Sola,
  Basilakos, and Plionis}}]{Grande:2011xf}
\bibinfo{author}{\bibfnamefont{J.}~\bibnamefont{Grande}},
  \bibinfo{author}{\bibfnamefont{J.}~\bibnamefont{Sola}},
  \bibinfo{author}{\bibfnamefont{S.}~\bibnamefont{Basilakos}},
  \bibnamefont{and} \bibinfo{author}{\bibfnamefont{M.}~\bibnamefont{Plionis}},
  \bibinfo{journal}{JCAP} \textbf{\bibinfo{volume}{08}}, \bibinfo{pages}{007}
  (\bibinfo{year}{2011}), \eprint{1103.4632}.

\bibitem[{\citenamefont{Basilakos et~al.}(2012)\citenamefont{Basilakos,
  Polarski, and Sola}}]{Basilakos:2012ra}
\bibinfo{author}{\bibfnamefont{S.}~\bibnamefont{Basilakos}},
  \bibinfo{author}{\bibfnamefont{D.}~\bibnamefont{Polarski}}, \bibnamefont{and}
  \bibinfo{author}{\bibfnamefont{J.}~\bibnamefont{Sola}},
  \bibinfo{journal}{Phys. Rev. D} \textbf{\bibinfo{volume}{86}},
  \bibinfo{pages}{043010} (\bibinfo{year}{2012}), \eprint{1204.4806}.

\bibitem[{\citenamefont{Lima et~al.}(2013)\citenamefont{Lima, Basilakos, and
  Sola}}]{Lima:2013dmf}
\bibinfo{author}{\bibfnamefont{J.~A.~S.} \bibnamefont{Lima}},
  \bibinfo{author}{\bibfnamefont{S.}~\bibnamefont{Basilakos}},
  \bibnamefont{and} \bibinfo{author}{\bibfnamefont{J.}~\bibnamefont{Sola}},
  \bibinfo{journal}{Mon. Not. Roy. Astron. Soc.}
  \textbf{\bibinfo{volume}{431}}, \bibinfo{pages}{923} (\bibinfo{year}{2013}),
  \eprint{1209.2802}.

\bibitem[{\citenamefont{Sola}(2013)}]{Sola:2013gha}
\bibinfo{author}{\bibfnamefont{J.}~\bibnamefont{Sola}}, \bibinfo{journal}{J.
  Phys. Conf. Ser.} \textbf{\bibinfo{volume}{453}}, \bibinfo{pages}{012015}
  (\bibinfo{year}{2013}), \eprint{1306.1527}.

\bibitem[{\citenamefont{Solà and Gómez-Valent}(2015)}]{Sola:2015rra}
\bibinfo{author}{\bibfnamefont{J.}~\bibnamefont{Solà}} \bibnamefont{and}
  \bibinfo{author}{\bibfnamefont{A.}~\bibnamefont{Gómez-Valent}},
  \bibinfo{journal}{Int. J. Mod. Phys.} \textbf{\bibinfo{volume}{D24}},
  \bibinfo{pages}{1541003} (\bibinfo{year}{2015}), \eprint{1501.03832}.

\bibitem[{\citenamefont{Solà et~al.}(2017)\citenamefont{Solà, Gómez-Valent,
  and de~Cruz~Pérez}}]{Sola:2016jky}
\bibinfo{author}{\bibfnamefont{J.}~\bibnamefont{Solà}},
  \bibinfo{author}{\bibfnamefont{A.}~\bibnamefont{Gómez-Valent}},
  \bibnamefont{and}
  \bibinfo{author}{\bibfnamefont{J.}~\bibnamefont{de~Cruz~Pérez}},
  \bibinfo{journal}{Astrophys. J.} \textbf{\bibinfo{volume}{836}},
  \bibinfo{pages}{43} (\bibinfo{year}{2017}), \eprint{1602.02103}.

\bibitem[{\citenamefont{Wilson-Ewing}(2012)}]{Wilson-Ewing:2011gnq}
\bibinfo{author}{\bibfnamefont{E.}~\bibnamefont{Wilson-Ewing}},
  \bibinfo{journal}{Class. Quant. Grav.} \textbf{\bibinfo{volume}{29}},
  \bibinfo{pages}{085005} (\bibinfo{year}{2012}), \eprint{1108.6265}.

\bibitem[{\citenamefont{Cailleteau et~al.}(2012)\citenamefont{Cailleteau,
  Barrau, Grain, and Vidotto}}]{Cailleteau:2012fy}
\bibinfo{author}{\bibfnamefont{T.}~\bibnamefont{Cailleteau}},
  \bibinfo{author}{\bibfnamefont{A.}~\bibnamefont{Barrau}},
  \bibinfo{author}{\bibfnamefont{J.}~\bibnamefont{Grain}}, \bibnamefont{and}
  \bibinfo{author}{\bibfnamefont{F.}~\bibnamefont{Vidotto}},
  \bibinfo{journal}{Phys. Rev. D} \textbf{\bibinfo{volume}{86}},
  \bibinfo{pages}{087301} (\bibinfo{year}{2012}), \eprint{1206.6736}.

\end{thebibliography}
\end{document}